\documentclass[iop]{emulateapj}

\newcommand{\src}{\mbox{V1723~Aql}}

\begin{document}

\title{EVLA Nova Project Observations of the Classical Nova V1723~Aquilae}

\author{Miriam I. Krauss, Laura Chomiuk\altaffilmark{1, 2}, Michael Rupen, Nirupam Roy\altaffilmark{2}, and Amy J. Mioduszewski}
\affil{National Radio Astronomy Observatory,
Socorro, NM 87801 \\
mkrauss, lchomiuk, mrupen, amiodusz, and nroy@nrao.edu}
\altaffiltext{1}{Also Harvard-Smithsonian Astrophysical Observatory, Cambridge, MA 02139}
\altaffiltext{2}{National Radio Astronomy Observatory Jansky Fellow}

\author{J. L. Sokoloski}
\affil{Columbia Astrophysics Laboratory, Columbia University, New York, NY 10027 \\
jsokoloski@astro.columbia.edu}

\author{Thomas Nelson\altaffilmark{3} and Koji Mukai}
\affil{CRESST and X-ray Astrophysics Laboratory, NASA/GSFC, Greenbelt, MD 20771 \\
and Center for Space Science and Technology, University of Maryland Baltimore County, Baltimore, MD 21250 \\
thomas.nelson and koji.mukai-1@nasa.gov}
\altaffiltext{3}{Current address: School of Physics and Astronomy, University of Minnesota, Minneapolis, MN 55455}

\author{M. F. Bode}
\affil{Astrophysics Research Institute, Liverpool John Moores University, Twelve Quays House, Egerton Wharf, Birkenhead CH41 1LD \\
mfb@astro.livjm.ac.uk}

\author{S. P. S. Eyres}
\affil{Jeremiah Horrocks Institute, University of Central Lancashire, Preston, PR1 2HE, UK \\
spseyres@uclan.ac.uk}

\author{T. J. OÕBrien}
\affil{Jodrell Bank Centre for Astrophysics, University of Manchester, Manchester, UK M13 9PL \\ 
tim.obrien@manchester.ac.uk}


\begin{abstract}

We present radio light curves and spectra of the classical nova \src\ obtained with the Expanded Very Large Array (EVLA).  This is the first paper to showcase results from the EVLA Nova Project, which comprises a team of observers and theorists utilizing the greatly enhanced sensitivity and frequency coverage of EVLA radio observations, along with observations at other wavelengths, to reach a deeper understanding of the energetics, morphology, and temporal characteristics of nova explosions.  Our observations of \src\ span 1--37~GHz in frequency, and we report on data from 14--175~days following the time of the nova explosion.  The broad frequency coverage and frequent monitoring show that the radio behavior of \src\ does not follow the classic Hubble-flow model of homologous spherically expanding thermal ejecta.  The spectra are always at least partially optically thin, and the flux rises on faster timescales than can be reproduced with linear expansion.  Therefore, any description of the underlying physical processes must go beyond this simple picture.  The unusual spectral properties and light curve evolution might be explained by multiple emitting regions or shocked material.   Indeed, X-ray observations from {\it Swift} reveal that shocks are likely present.

\end{abstract}

\section{Introduction}
\begin{deluxetable*}{rlrccccccc}
\tablewidth{0pt}
\tabletypesize{\scriptsize}
\tablecaption{Observed 1--8~GHz Radio Flux Densities (mJy)\tablenotemark{a}}
\tablehead{
\colhead{} & \colhead{} & \colhead{} & \colhead{} & \multicolumn{6}{c}{Frequency (GHz)} \\
\cline{5-10}
\colhead{Date} & \colhead{MJD} & \colhead{Day\tablenotemark{a}} & \colhead{Epoch} & \colhead{1.38} & \colhead{1.88} & \colhead{4.74} & \colhead{5.25} & \colhead{6.49} & \colhead{7.00}}
\startdata
25 Sep 2010 & 55464.1 &  13.6 &  1 & \nodata & \nodata & $<0.042$\tablenotemark{c} & \nodata & $<0.033$\tablenotemark{c} & \nodata \\
6 Oct 2010 & 55475.1 &  24.6 &  3 & \nodata & \nodata & 0.096$\pm$0.025 & 0.122$\pm$0.025 & 0.155$\pm$0.020 & 0.161$\pm$0.021 \\
15 Oct 2010 & 55484.1 &  33.6 &  4 & \nodata & \nodata & 0.536$\pm$0.029 & 0.606$\pm$0.028 & 0.778$\pm$0.023 & 0.857$\pm$0.023 \\
18 Oct 2010 & 55487.1 &  36.6 &  5 & $<0.38$ & $<0.32$ & 0.964$\pm$0.045 & 1.030$\pm$0.039 & 1.279$\pm$0.036 & 1.393$\pm$0.039 \\
24 Oct 2010 & 55493.1 &  42.6 &  6 & $<0.65$ & 1.26$\pm$0.11 & 2.194$\pm$0.042 & 2.390$\pm$0.041 & 2.892$\pm$0.040 & 3.044$\pm$0.041 \\
29 Oct 2010 & 55498.0 &  47.5 &  7 & $<0.81$ & 1.39$\pm$0.18 & 3.193$\pm$0.050 & 3.490$\pm$0.051 & 4.072$\pm$0.051 & 4.285$\pm$0.050 \\
7 Nov 2010 & 55507.8 &  57.3 &  8 & $<1.61$ & 1.30$\pm$0.20 & 3.910$\pm$0.052 & 4.079$\pm$0.051 & 4.572$\pm$0.054 & 4.791$\pm$0.054 \\
2 Dec 2010 & 55532.9 &  82.4 &  9 & $<0.84$ & 0.78$\pm$0.18 & 1.368$\pm$0.030 & 1.393$\pm$0.028 & 1.922$\pm$0.034 & 2.048$\pm$0.028 \\
18 Dec 2010 & 55548.9 &  98.4 & 10 & $<0.81$ & $<0.57$ & 1.124$\pm$0.038 & 1.256$\pm$0.034 & 1.559$\pm$0.032 & 1.677$\pm$0.036 \\
14 Jan 2011 & 55575.8 & 125.3 & 11 & \nodata & \nodata & 0.996$\pm$0.038 & 1.133$\pm$0.037 & 1.389$\pm$0.034 & 1.497$\pm$0.032 \\
30 Jan 2011 & 55591.8 & 141.3 & 12 & $<0.75$ & $<0.43$ & 0.885$\pm$0.055 & 0.967$\pm$0.052 & 1.225$\pm$0.051 & 1.351$\pm$0.046 \\
5 Mar 2011 & 55625.5 & 175.0 & 13 & \nodata & \nodata & 0.669$\pm$0.021 & 0.787$\pm$0.019 & 1.159$\pm$0.023 & 1.201$\pm$0.020 \\
\enddata
\label{srcTab1}
\tablenotetext{a}{Quoted upper limits are $3\sigma$.}
\tablenotetext{b}{Days from 11 September 2010 (MJD 55450.5).}
\tablenotetext{c}{Insufficient source flux to perform self-calibration.}
\end{deluxetable*}

Novae are the most common major stellar explosions in the Universe, resulting when the
accretion of hydrogen onto the surface of a white dwarf leads to a thermonuclear runaway. 
Galactic novae occur relatively nearby and evolve on short timescales (months to years), making them
excellent laboratories for understanding the accretion and outflow processes that influence a wide
range of astrophysical phenomena, from planetary nebulae to supernovae and active galactic nuclei.
Radio continuum emission from classical novae has been observed for more than three decades
\citep{hjellming79,seaquist77,seaquist80}, and is generally thought to result from thermal free-free processes.  Most novae observed in the radio have been understood as expanding, thermally-emitting shells of ejecta, though some show evidence of non-thermal emission as well (see \citealt{seaquist08} for a review).   Radio continuum emission provides unique insights into the
properties of this ejecta, since it is optically thick at much lower densities than other wavelengths (e.g.,
optical emission), is not subject to extinction, and can 
provide a measure of the total ejecta mass, temperature and density profiles, and kinetic energy \citep{hjellming96}.  

The EVLA Nova Project, undertaken by a team including observers and theorists of radio, optical, and X-ray emission from novae, is embarking on a new era of observations of Galactic novae.  The cornerstone of the EVLA Nova Project is EVLA light curves and spectra of unprecedented time resolution, frequency coverage, and sensitivity.  We endeavor to monitor every new nova in the Galaxy which is closer than 5~kpc, producing high-quality radio light curves and spectra which allow us to compare, in exquisite detail, the radio properties of novae with current theory, and will also provide the data needed to go beyond the simple, standard models of radio emission from novae.  In addition, we are utilizing information from high-energy observations, as well as very-long-baseline radio interferometry, to help synthesize a complete, multifrequency understanding of nova explosions.

Here, we present observations of the classical nova \src, which was discovered in outburst on 2010 September 11; we take the time of the start of the outburst to be MJD
55450.5  \citep[][]{yamanaka10,balam10}. It is highly
reddened, and on the first day of the outburst the $\rm H_\alpha$ emission
line had a full-width at zero intensity of 3000~km~s$^{-1}$ \citep{yamanaka10}, from which we infer an expansion velocity for the nova ejecta of 1500~km~s$^{-1}$. From a peak R-band magnitude of 13.43 on 11.6 September 2010 \citep{yamanaka10}, it took 20 days for the R-band flux to decrease by two magnitudes \citep{henden10}, implying that V1723 Aql is a member of the ``fast'' nova speed class \citep{payne57}.

We describe our EVLA radio and {\it Swift} X-ray observations and data reduction procedures in \S\ref{obsSec} and data analysis in \S\ref{analysis}.  We discuss possible physical interpretations of our observations in \S\ref{discSec}.

\section{Observations}\label{obsSec}

\subsection{EVLA Observations}

We monitored \src\ with the EVLA beginning on 2010 September 25, approximately two weeks after its initial optical discovery.  Here, we present data from observations taken through 2011 March 5; further observations are planned and will be published separately.  Each monitoring epoch covers a wide range of frequencies; most epochs include data
spanning 1--37~GHz.  During the course of our monitoring, which comprises a total of 27 hours and 840 GB of raw data, the EVLA progressed through the DnC, C, CnB, and B configurations.  All data were acquired with the widest possible bandwidths using the WIDAR correlator  and the current 8-bit samplers: 2 GHz at each band except L and X, where we used the available 1 and 0.8 GHz of bandwidth.

For each band, we observed a standard flux-reference calibration source, either 3C48 or 3C286, as well as a phase reference calibrator, either J1822-0938 (8.6\degr~away; 1--9~GHz) or J1851+0035 (4.5\degr~away; 19--37~GHz).  Processing was done with NRAO's Common Astronomy Software Applications (CASA) and Astronomical Image Processing System (AIPS).  For each package, the analysis path was the same: first, we obtained a bandpass solution using a bright calibration source (usually the flux calibrator).  With this solution, we solved for the gain amplitude and phase, and scaled the phase calibration source's amplitude from the flux calibration source's values, using the ``Perley-Butler 2010'' flux density standard. Finally, we applied these gain solutions to \src.  We performed one round of phase-only self-calibration using a point source model at 19--37~GHz; at lower frequencies (since other nearby sources are present) we used an image of the field for reference and two to three rounds as needed for convergence.  Errors on the flux include errors from gaussian fitting, the image rms noise, and 
 systematic errors of 1\% at low frequencies (1--9~GHz) and 3\% at high frequencies (19--37~GHz).   Since these data were taken during the commissioning phase of the EVLA, our estimates of the systematics are based on the apparent scatter in the broad-band spectra.  These errors are included in all tables and figures.  For analysis, we grouped our data into ``epochs'', with observations included in a given epoch separated by no more than three days.  Tables~\ref{srcTab1} and~\ref{srcTab2} list the flux densities at each epoch. 

\begin{deluxetable*}{rlrccccccc}
\tablewidth{0pt}
\tabletypesize{\scriptsize}
\tablecaption{Observed 8--37~GHz Radio Flux Densities (mJy)\tablenotemark{a}}
\tablehead{
\colhead{} & \colhead{} & \colhead{} & \colhead{} & \multicolumn{6}{c}{Frequency (GHz)} \\
\cline{5-10}
\colhead{Date} & \colhead{MJD} & \colhead{Day\tablenotemark{a}} & \colhead{Epoch} & \colhead{8.24} & \colhead{8.73} & \colhead{20.1} & \colhead{25.6} & \colhead{28.2} & \colhead{36.5}}
\startdata
25 Sep 2010 & 55464.1 &  13.6 &  1 & \nodata & \nodata & \nodata & \nodata & $<  0.09$\tablenotemark{d} & \nodata \\
3 Oct 2010 & 55472.1 &  21.6 &  2 & \nodata & \nodata & \nodata & \nodata &   0.43$\pm$ 0.07\tablenotemark{c,d} & \nodata \\
6 Oct 2010 & 55475.1 &  24.6 &  3 & \nodata & \nodata & \nodata & \nodata &   0.90$\pm$ 0.10\tablenotemark{d} & \nodata \\
15 Oct 2010 & 55483.1 &  32.6 &  4 & \nodata & \nodata & \nodata & \nodata &   2.46$\pm$ 0.11\tablenotemark{d} & \nodata \\
18 Oct 2010 & 55487.1 &  36.6 &  5 &  1.604$\pm$0.042 &  1.741$\pm$0.055 &   \nodata &   \nodata &   \nodata &   \nodata \\
19 Oct 2010 & 55488.1 &  37.6 &  5 & \nodata &  \nodata &   2.84$\pm$ 0.09 &   3.89$\pm$ 0.12 &   4.22$\pm$ 0.14 &   4.64$\pm$ 0.15 \\
24 Oct 2010 & 55493.1 &  42.6 &  6 &  3.492$\pm$0.052 &  3.648$\pm$0.055 & \nodata & \nodata & \nodata & \nodata \\
29 Oct 2010 & 55497.0 &  46.5 &  7 &  4.817$\pm$0.060 &  4.891$\pm$0.063 &   7.02$\pm$ 0.21 &   7.87$\pm$ 0.24 &   7.63$\pm$ 0.23 &   8.94$\pm$ 0.28 \\
7 Nov 2010 & 55507.8 &  57.3 &  8 &  5.299$\pm$0.064 &  5.393$\pm$0.067 &   7.20$\pm$ 0.23 &   8.12$\pm$ 0.25 &   8.72$\pm$ 0.27 &   9.20$\pm$ 0.28 \\
3 Dec 2010 & 55533.9 &  83.4 &  9 &  2.490$\pm$0.039 &  2.554$\pm$0.045 &   5.83$\pm$ 0.18 &   7.07$\pm$ 0.22 &   8.00$\pm$ 0.25 &   9.56$\pm$ 0.31 \\
18 Dec 2010 & 55548.9 & 98.4 & 10 &  1.938$\pm$0.038 &  2.010$\pm$0.045 &   \nodata &   \nodata &   \nodata &  \nodata \\
21 Dec 2010 & 55551.7 & 101.2 & 10 &  \nodata &  \nodata &   5.45$\pm$ 0.16 &   7.09$\pm$ 0.21 &   8.34$\pm$ 0.26 &  10.98$\pm$ 0.34 \\
14 Jan 2011 & 55575.8 & 125.3 & 11 &  1.676$\pm$0.037 &  1.780$\pm$0.044 &   4.99$\pm$ 0.15 &   6.09$\pm$ 0.19 &   6.53$\pm$ 0.20 &   8.51$\pm$ 0.27 \\
30 Jan 2011 & 55591.8 & 141.3 & 12 &  1.620$\pm$0.035 &  1.703$\pm$0.040 &   4.53$\pm$ 0.14 &   6.22$\pm$ 0.19 &   6.95$\pm$ 0.22 &   8.91$\pm$ 0.28 \\
5 Mar 2011 & 55625.5 & 175.0 & 13 &  1.436$\pm$0.023 &  1.510$\pm$0.023 &   5.01$\pm$ 0.15 &   7.07$\pm$ 0.21 &   7.95$\pm$ 0.24 &  11.05$\pm$ 0.34 \\
\enddata
\label{srcTab2}
\tablenotetext{a}{Quoted upper limits are $3\sigma$.}
\tablenotetext{b}{Days from 11 September 2010 (MJD 55450.5).}
\tablenotetext{c}{Insufficient source flux to perform self-calibration.}
\tablenotetext{d}{Values are reported for 32.1 GHz.}
\end{deluxetable*}

\subsection{{\it Swift} Observations}

On days 40.6 and 61.4, we obtained observations of \src\ with the {\it Swift} satellite \citep{Gehrels04} of duration 3.0 and 5.5 ks.  Each observation resulted in an exposure with the X-ray telescope (XRT) and an image with the Ultraviolet Optical Telescope (UVOT).  We used the {\it Swift} data analysis routines in HEASoft  version 6.10 throughout our analysis.  The UVOT images were both obtained with the UVW1 filter, which has a central wavelength of 2600 \AA~and a FWHM of 693 \AA~\citep{Poole08}.  \src\ was not detected in either UVOT image with a limiting UVW1 magnitude of 20.36 (20.65) in the first (second) observation.  

The XRT was operated in photon counting mode during both observations.    We produced cleaned level 2 event files by running the XRT reduction pipeline on the level 1 event files, retaining events with grades 0--12.  We then used XSelect v2.4 to create spectra for each dataset.  Source counts were extracted from 30\arcsec-radius circular region centered on \src; background counts were extracted from a 95\arcsec-radius source free region.  The ancilliary response files (ARF) were generated using the task {\tt xrtmkarf}, and were corrected for hot pixels and dead columns using an exposure map of each observation.  Finally, we used the most recent response matrix file (RMF) appropriate for PC mode and event grades 0--12 from the {\it Swift} calibration database.    

\section{Data Analysis and Results}\label{analysis}

\begin{figure}
\vspace{-0.3in}
\resizebox{\columnwidth}{!}{\includegraphics{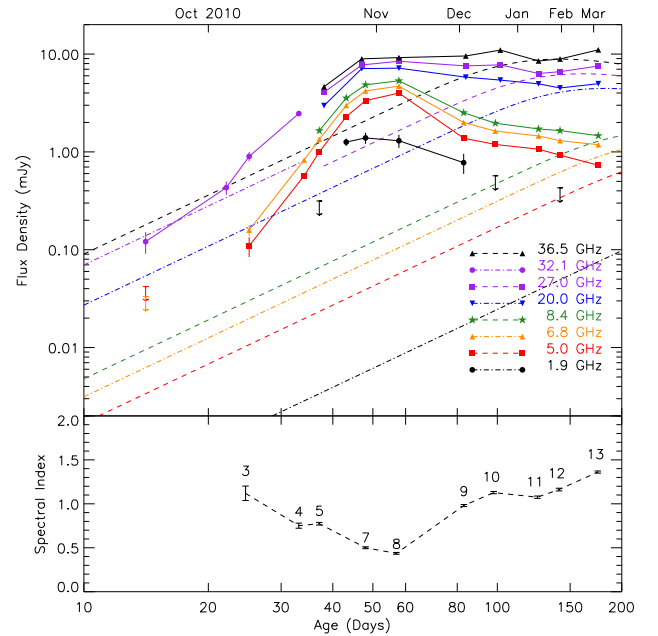}}
	\caption{{\bf Upper panel}: EVLA light curves of \src\ (solid lines; $3\sigma$ upper limits are plotted as downward arrows).  Dashed and dot-dashed curves show Hubble-flow model light curves for $M_{\rm ej} = 10^{-4}~M_{\sun}$, $T_{\rm ej} = 10^4 \rm~K$, $v_{\rm ej} = 1500 \rm~km~s^{-1}$, and $d = 5$~kpc.  Decreasing the distance will raise all model light curves equally in log space, so will not change their spread in frequency or their poor match to the observed fluxes.  This discrepancy is due to the fact that the observed spectral index during this time is significantly less than $\alpha=2$ ($S_{\nu} \propto \nu^{\alpha}$), which is the value predicted by the optically-thick rise of the Hubble-flow model.  {\bf Lower panel}: the spectral index $\alpha$ from a single power-law fit to the 4--37~GHz data; each point is labeled by observing epoch.}
	\label{lcPlot}
\end{figure}

\begin{deluxetable*}{cccccccc}
\vspace{-0.3in}
\tablewidth{0pt}
\tabletypesize{\small}
\tablecaption{Power-law Spectral Indices of Radio Spectra}
\tablehead{
\colhead{} & \multicolumn{2}{c}{4--40 GHz} & \multicolumn{2}{c}{4--9 GHz} & \multicolumn{2}{c}{19--40 GHz} & \colhead{F-test\tablenotemark{a}} \\
\colhead{Epoch} & \colhead{$\alpha$} & \colhead{$\chi^2_{\nu}$ (dof)\tablenotemark{b}}  & \colhead{$\alpha$} & \colhead{$\chi^2_{\nu}$ (dof)} & \colhead{$\alpha$} & \colhead{$\chi^2_{\nu}$ (dof)} & \colhead{$\sigma$}} 
\startdata
 3 & $ 1.120\pm 0.082$ &  0.07 (3) & \nodata & \nodata & \nodata & \nodata & \nodata \\
 4 & $ 0.751\pm 0.029$ &  4.23 (3) & \nodata & \nodata & \nodata & \nodata & \nodata \\
 5 & $ 0.774\pm 0.015$ &  2.44 (8) & $ 0.978\pm 0.067$ &  0.20 (4) & $ 0.778\pm 0.070$ &  4.52 (2) & 0.99 \\
 6 & \nodata & \nodata &  $ 0.840\pm 0.031$ &  0.15 (4) & \nodata & \nodata & \nodata \\
 7 & $ 0.499\pm 0.011$ & 10.77 (8) & $ 0.710\pm 0.026$ &  0.67 (4) & $ 0.363\pm 0.068$ &  1.53 (2) & 2.86 \\
 8 & $ 0.436\pm 0.010$ &  4.13 (8) & $ 0.551\pm 0.024$ &  0.40 (4) & $ 0.418\pm 0.073$ &  0.95 (2) & 2.46 \\
 9 & $ 0.980\pm 0.012$ &  4.93 (8) & $ 1.133\pm 0.036$ &  3.81 (4) & $ 0.843\pm 0.074$ &  0.56 (2) & 1.17 \\
10 & $ 1.127\pm 0.013$ &  1.98 (8) & $ 0.956\pm 0.049$ &  0.26 (4) & $ 1.188\pm 0.072$ &  0.62 (2) & 2.18 \\
11 & $ 1.075\pm 0.013$ &  2.46 (8) & $ 0.912\pm 0.056$ &  0.60 (4) & $ 0.891\pm 0.073$ &  0.41 (2) & 2.05 \\
12 & $ 1.162\pm 0.015$ &  0.34 (8) & $ 1.116\pm 0.077$ &  0.05 (4) & $ 1.134\pm 0.072$ &  0.98 (2) & 0.61 \\
13 & $ 1.361\pm 0.012$ &  4.03 (8) & $ 1.281\pm 0.041$ &  6.79 (4) & $ 1.321\pm 0.072$ &  0.21 (2) & 0.55 \\
\enddata
\label{radFits}
\tablenotetext{a}{The F-test tests the null hypothesis that the reduced $\chi^2$ values are the same for the single-- and two--power-law fits, accounting for two fewer degrees of freedom.}
\tablenotetext{b}{Reduced $\chi^2$ (degrees of freedom).}
\end{deluxetable*}

\subsection{Radio Light Curves and Spectra}\label{radSec}

\begin{figure}
\resizebox{\columnwidth}{!}{\includegraphics{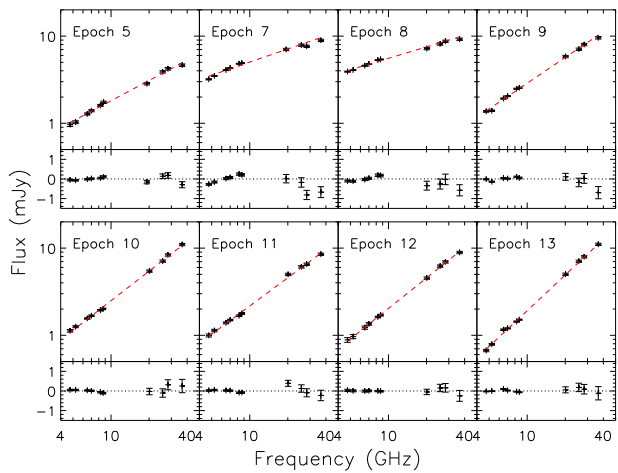}}
	\caption{Spectral fits and fit residuals for each epoch where there are flux measurements for 4--37~GHz.  Red dashed lines show the single power-law fits for this entire frequency range (residuals are plotted below each spectrum).} 
	\label{specPlot}
\end{figure}

\src's light curves are characterized by an initial steep rise until around day 50  ($S_{\nu} \propto t^{3.3}$ at 32.1~GHz), at which point there is a turnover and rapid decay at the lower frequencies (1--9~GHz) and flattening at 19~GHz and higher (Figure~\ref{lcPlot}).  In order to characterize the radio spectra, we fit simple power-law models to each epoch's 4--37~GHz spectrum; the spectra, along with fits and residuals, are shown in Figure~\ref{specPlot}.  The fit residuals in some epochs suggest the presence of a spectral break, so we performed separate power-law fits to the lower (1--9~GHz) and higher (19--37~GHz) frequencies as well.  In order to determine whether any improvement in fit was statistically significant, we performed F-tests for each epoch.  The addition of a second power-law is favored at less than $3\sigma$ significance, so we use the results from the single power-law fits in our discussion of \src.  We also plot this spectral index $\alpha$ ($S_{\nu} \propto \nu^{\alpha}$) in the lower panel of Figure~\ref{lcPlot}, showing that the rise portion of the light curve is accompanied by a flattening of the spectrum, which then steepens again with the turnover in low-frequency flux. 
We report the results of our spectral fitting in Table~\ref{radFits}.  

\subsection{X-ray Modeling}

The {\it Swift} XRT detected 25 (47) events at the position of \src\ on day 41 (61).  We estimate 2 (4) background counts, resulting in count rates of $7.6\pm1.6$ ($7.8\pm1.2$) $\times \,10^{-3}$~c~s$^{-1}$.   We fit the unbinned, non-background subtracted X-ray data with absorbed power-law and thermal bremsstrahlung models in the XSpec spectral fitting package (version 12.6.0) using the Cash statistic \citep{Cash79}.  Although the limited number of counts precludes the determination of well-constrained limits on the fit parameters, the lack of photons with $E < 1\rm~keV$ in the first observation, and the appearance of such photons 
in the second observation, suggests a decreasing absorbing column.  The power-law fits find a spectral index of  
$-2.0^{+2.1}_{-2.2}$ ($-1.4^{+0.9}_{-1.1}$), absorbing column of $11^{+11}_{-8}$ ($1.3^{+1.9}_{-1.0}$) $\times \, 10^{22}\,\rm cm^{-2}$, and flux of $1.3^{+0.4}_{-1.3}$ ($0.9^{+0.3}_{-0.9}$) $\times \, 10^{-12}$ erg cm$^{-2}$ s$^{-1}$ for the first (second) observation.  The thermal bremsstrahlung fits yield similar values for the absorbing column and flux, and place limits on the temperature of the X-ray emitting material in the first (second) observation of $T > 2.1$ (3.4)  $\times \, 10^7$ K (90\% confidence). 

\section{Discussion}\label{discSec}

A standard model for radio emission from novae envisions a thick, spherical shell of warm ($T \sim 10^4$~K), thermal-bremsstrahlung--emitting ejecta moving away from the white dwarf following the initial explosion \citep{hjellming79}.  This is often known as the ``Hubble-flow'' model, since the velocity of the ejecta scales linearly with radial distance from the white dwarf ($v \propto r$).  During the rise portion of the nova light curve, the ejecta are optically thick, while the  radio photosphere is approximately coincident with the outer shell boundary.
This yields a spectrum with $S_{\nu} \propto \nu^2$ and flux that increases with emitting area  ($S_{\nu} \propto t^2$).  
 As the shell expands, the density falls and the free-free opacity drops, first at high and then at lower frequencies.  The source transitions to an optically thin spectrum ($\alpha = -0.1$), with the ``width" of the transition, in time or in frequency, proportional to the physical width of the emitting shell.

This model has worked very well for a number of novae, including some that have been imaged with MERLIN and the VLA \citep[e.g., V1500~Cyg, QU~Vul, and V723~Cas;][]{hjellming79,taylor88,heywood05}.  Note that overall shape of  light curves determined by the Hubble-flow model is fixed by a small set of parameters: the distance~$d$, ejected mass~$M_{\rm ej}$, ejecta temperature~$T_{\rm ej}$, outer ejecta velocity~$v_{\rm ej}$, and ratio of inner to outer velocity (which is not relevant until late in the light curve evolution).  During the rise portion of the light curve, no modification of these parameters will change the fact that the Hubble-flow model predicts an optically thick ($\alpha = 2$) source which increases in flux as $t^2$.  Neither of these basic characteristics are observed in \src.  To characterize this discrepancy, we compare our light curves with a set calculated using the Hubble-flow model.  We use characteristic values for classical novae of $M_{\rm ej} = 10^{-4}~M_{\sun}$ and $T_{\rm ej} = 10^4 \rm~K$ \citep[e.g., Table~7.2 of][]{seaquist08}, as well as the outer ejecta velocity from optical line widths, $v_{\rm ej} = 1500 \rm~km~s^{-1}$. 
So that these light curves do not, at any point, overpredict the observed fluxes, we assume a distance of $d = 5$~kpc.  If there is an underlying Hubble-flow component, this assumption allows for its presence in addition to other possible contributions to the flux.  Furthermore, changes in the assumed distance, which is otherwise ill-defined for this nova, will simply scale the flux and will not change the shape of the light curves or the spectral properties.

The observed light curves show a broad ``bump'' of excess emission relative to these model light curves (Figure~\ref{lcPlot}).
The initial rise goes as $t^{3.3}$---much faster than the expected $t^2$ 
    from the freely-expanding shell of the Hubble-flow model---while the initial spectrum
    has $\alpha = {1.1}$, substantially shallower than the $\alpha = 2$ expected from its
    purely optically thick, single-temperature emission.
 During the time when the bump is present, the spectral index decreases significantly, reaching a minimum of $\alpha \approx 0.4$.  The shallow spectral index implies that the source is partially optically thin during the entire bump, over a factor of $\sim100$ change
    in flux density.  Most astronomical transients, whether supernovae,
    quasars, X-ray binaries, or  gamma-ray bursts, show rapid increases in
    flux density only when they are optically thick, and peak when they
    turn optically thin.
Finally, the data during the decline show the spectrum {\it steepening},
    with the flux densities at the higher frequencies remaining almost
    constant while the flux densities at the lower frequencies drop dramatically.
    By the time of our last observation, the spectral index had risen to
    $\alpha = 1.4$.  This could be the appearance of the
    expected Hubble-flow emission. However, the relaxation to a more optically thick
    spectrum is the opposite of normal behavior from expanding ejecta, where a source
    becomes more optically thin with time.

This radio behavior --- particularly the rapid rise while optically thin --- is,
to our knowledge, unique among both classical novae and astronomical
transients in general.  The limited optical and X-ray data show 
\src\ as a normal, unremarkable classical nova.  The radio oddities seen here would have been easily visible in well-sampled light-curves of novae such as V1974~Cyg or FH~Ser~1970 \citep{hjellming96}, 
assuming they were of a similar magnitude relative to the observed optically-thick emission.

The radio data do not obviously suggest any emission process beyond thermal
bremsstrahlung.  Although synchrotron emission can, in principle, produce a wide range of
spectral indices, it is commonly seen with a much steeper rise when optically thick
($\alpha = 2.5$) or a falling spectrum when optically thin ($\alpha =-0.5$ to
$-1$).  Furthermore, there is no obvious signature of a dense environment which might lead
to strong shocks to power the
synchrotron emission \citep[as in RS Oph;][]{obrien06,rupen08, sokoloski08}. We do not see evidence for excess emission at low frequencies
($\sim 1$--$2$ GHz), and can place a 3$\sigma$ upper limit on the linear polarization
of $<$39 $\mu$Jy (or 1\%) on day 48, averaging across 1~GHz of bandwidth from 5--6~GHz, 
and less than 0.36 mJy (or 10\%), averaging
across a single channel of 2 MHz bandwidth.
However, these non-detections are not surprising, considering
the high levels of Faraday depolarization expected in the Galactic plane, as well as beam depolarization and depolarization internal to the nova itself. 

Assuming the radio emission process is thermal bremsstrahlung, we must make
significant revisions to the standard Hubble-flow model to explain our
observations.  The simplest possibility is that there are two physically
distinct emission regions, one giving rise to the bump and the other to
the late-time, optically thick emission.  The challenge here is that the
optically thick material cannot obscure the bump emission. Either the
source is highly asymmetric, with the bump material physically distinct
from the optically thick region (e.g., a bipolar outflow plus an expanding
thick ring), or the material producing the bump must lie outside the
optically thick region.

If there are physically distinct emitting regions, the optically thick emission might correspond to
the early stages (rising flux density part) of a modified Hubble-flow model.
The bump is more difficult to explain, since this requires increasing
emission from optically {\it thin} components over time.  Thermal
bremsstrahlung requires hot ionized gas, and the rapid flux density rise
requires rapid physical evolution.  One obvious candidate for producing such
highly variable hot gas is shocks, either internal to the nova ejecta, 
or between the ejecta and surrounding circumstellar material
(CSM).  Both have been proposed for classical novae on other grounds.
The presence of internal shocks is motivated by the profiles and evolution 
of optical line spectra, which indicate that novae
eject material in two stages: an initial relatively slow ejection,
followed by a faster wind \citep[e.g.,][]{warner08}.  The material in the wind
eventually catches up with the initial ejecta, producing shocks with
characteristic velocity differentials of up to a few thousand km~s$^{-1}$.  Hard
X-ray emission ($>1$~keV) has also been detected from many classical novae,
5--1000 days after the optical peak (\citealt{mukai08} and references therein), 
and this has been modeled as arising from these same internal shocks
\citep{obrien94}.  Internal shocks have previously been posited
to explain early-time radio emission from Nova Vul 1984 No. 2 \citep{taylor87} 
as well.  There is less direct evidence for shocks between the nova
ejecta and a surrounding circumstellar medium, but such a CSM has been
proposed by \citet{williams10} and would naturally lead to shocks,
as are often seen in supernovae.

The {\it Swift} XRT detected \src\ as a hard ($E > 1$~keV) X-ray source during the time when
the bump was seen in the radio light curves, suggesting that strong shocks
were present.  While the X-ray--emitting gas itself would not produce
observable radio emission, hydrodynamical simulations of colliding outflows
suggest that the gas behind the shock might contain a range of temperatures
as well as regions of high density (\citealt{obrien94}; see also
\citealt{lloyd96}).  The required gas densities, temperatures, and total
masses are not excessive.  At 5~kpc, the observed 8\,GHz emission
around day 40 could be produced by a shell with thickness
$\sim10^{13}\rm\,cm$ (outer radius $\sim5\times10^{14}\rm\,cm$),
density $n_e\sim10^7\rm\,cm^{-3}$, temperature $\sim10^6\rm\,K$, and
total mass of a few times $10^{-6}\rm\,M_\sun$.  Whether such
models can reproduce the observed light curves and spectra remains to be seen. 
Any successful model must also explain why the bump seen in \src\ is not seen in the radio light curves of other classical novae, although many classical novae show evidence of shocks.
These models, and the relation
between \src\ and other novae, depend crucially on continued radio
monitoring, which will show whether \src\ ultimately follows the canonical
Hubble-flow models at late times.  Upcoming radio imaging with the EVLA in its most
extended, highest resolution ``A'' configuration will also provide valuable
information on the physical size and geometry of the ejecta.

\acknowledgments{The National Radio Astronomy Observatory is a facility of the National Science Foundation operated under cooperative agreement by Associated Universities, Inc.  We acknowledge with thanks the variable star observations from the AAVSO International database contributed by observers worldwide and used in this research.  This research has made use of NASA's Astrophysics Data System Bibliographic Services.}

\end{document}